\documentclass[aip,jcp,reprint]{revtex4-1}

\usepackage[usenames,dvipsnames]{color}
\usepackage{amsmath}
\usepackage{amssymb}
\usepackage{graphicx}
\usepackage{hyperref}

\usepackage{color}

\newcommand{\on}{{\overline{n}}}

\begin{document}

\title{ Free energy of ligand-receptor systems forming multimeric complexes}

\author{Lorenzo Di Michele}
\affiliation{Biological and Soft Systems, Cavendish Laboratory, University of Cambridge, JJ Thomson Avenue, Cambridge, CB3 0HE, United Kingdom}
\author{Stephan J.\ Bachmann}
\affiliation{Interdisciplinary Center for Nonlinear Phenomena and Complex Systems \&
Service de Physique des Syst\`{e}mes Complexes et M\'{e}canique Statistique, Universit\'{e} libre de Bruxelles (ULB), Campus Plaine, CP 231, Blvd du Triomphe, B-1050 Brussels, Belgium.}
\author{Lucia Parolini}
\affiliation{Biological and Soft Systems, Cavendish Laboratory, University of Cambridge, JJ Thomson Avenue, Cambridge, CB3 0HE, United Kingdom}
\author{Bortolo M. Mognetti }
\email{bmognett@ulb.ac.be}
\affiliation{Interdisciplinary Center for Nonlinear Phenomena and Complex Systems \&
Service de Physique des Syst\`{e}mes Complexes et M\'{e}canique Statistique, Universit\'{e} libre de Bruxelles (ULB), Campus Plaine, CP 231, Blvd du Triomphe, B-1050 Brussels, Belgium.}

\begin{abstract}
Ligand-receptor interactions are ubiquitous in biology and have become popular in materials in view of their applications to programmable self-assembly. Although, complex functionalities often emerge from the simultaneous interaction of more than just two linker molecules, state of art theoretical frameworks enable the calculation of the free energy only in systems featuring one-to-one ligand/receptor binding.  In this communication we derive a general formula to calculate the free energy { of  systems} 
featuring simultaneous direct interaction between an arbitrary number of linkers. To exemplify the potential and generality of our approach we apply it to the systems recently introduced by  Parolini {\em et al.} [\emph{ACS Nano}  \textbf{10}, 2392 (2016)] and Halverson {\em et al.} [\emph{J.\ Chem.\ Phys.}\ \textbf{144}, 094903 (2016)], both featuring functionalized Brownian particles interacting via three-linker complexes.
\end{abstract}

\maketitle
\begin{figure}
\includegraphics[width=7cm]{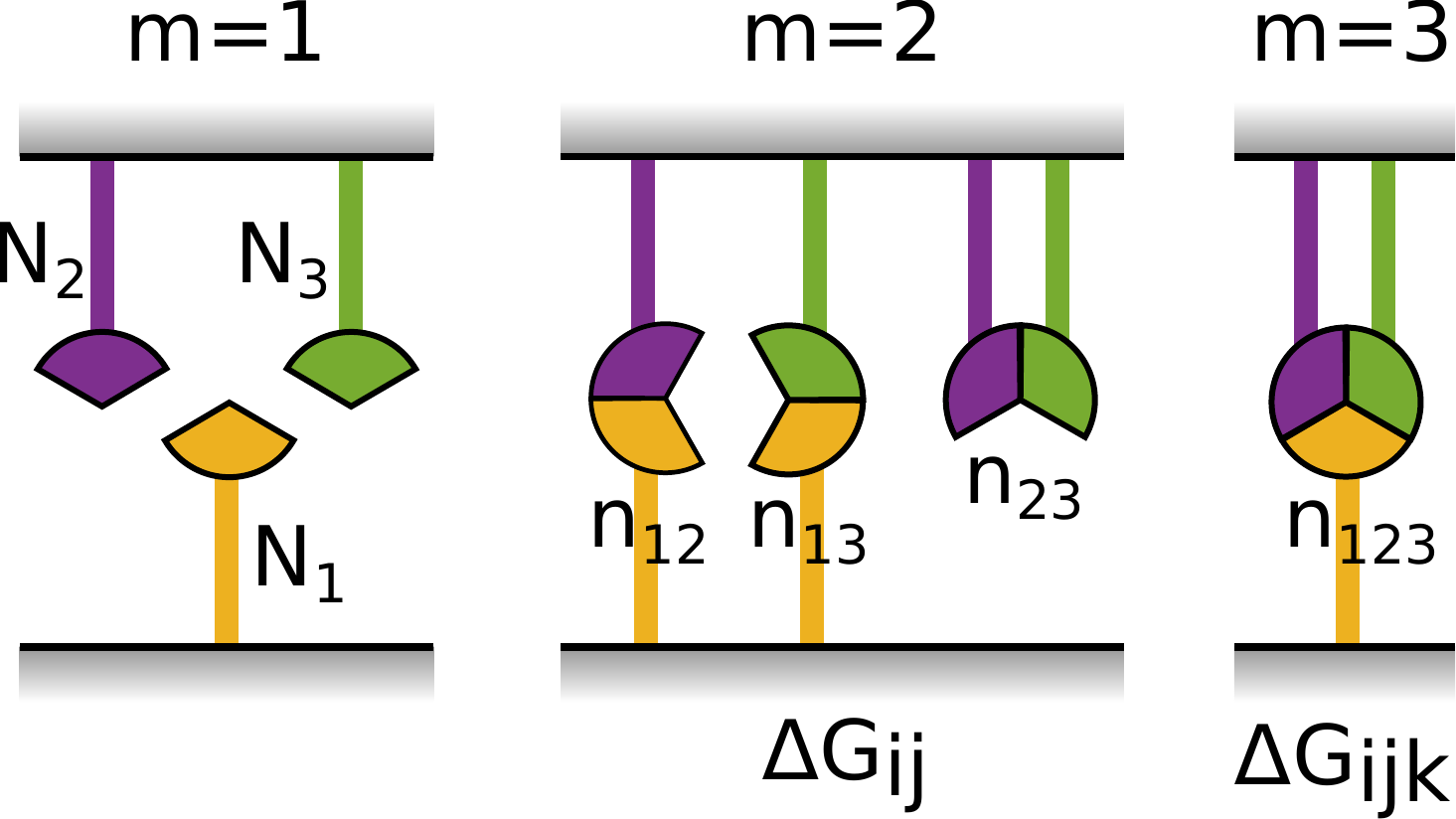} 
\caption{ A system of three families of linkers binding in pairs and { three-molecule} multimeric complexes \label{Fig1}}
\end{figure}

The quantitative understanding of the  ligand-receptor  interactions is receiving much attention  in view of the key role played in biology and their applications to the self-assembly of composite materials.\\
Biological cells respond to the presence of specific molecules via cell-surface receptors. Examples include \emph{toll-like receptors}, triggering immune response to bacterial and viral activity\cite{GayReview2014}, and \emph{receptor tyrosine kinases}, involved in the regulation of several physiological  processes \cite{lemmon2010cell}. In order for the signals to be transduced across the cell membrane, the presence of the ligands typically triggers dimerization or oligomerization of the receptors, through interactions that involve multiple molecules.\\
Functionalizing Brownian units with specific linkers, often made of synthetic DNA molecules, is a powerful tool to engineer the structure and response of self-assembled soft materials\cite{mirkin,alivisatos,crystal-mirkin2,lorenzo,parolini2014thermal,stefano-nature,rogers-manoharan,pine,wang2015crystallization}.
 Many functionalization schemes rely on one-to-one ligand-receptor interactions, but recently designs featuring multi-linker complexation have been proposed to extend the accessible range of functionalities\cite{Halverson2016,Parolini2016,romano2015switching,mcginley2013assembling,rogers-manoharan}.
In particular, Parolini {\em et al.\ }\cite{Parolini2016} adopted three-linker complexes enabling toehold-mediated strand exchange reactions\cite{zhang2009control} to control aggregation kinetics of lipid vesicles coated with DNA linkers.
%
%{\cm Multistrand interactions have also been used to control self-assembly of DNA--stars \cite{romano2015switching} and to design complex collective behaviours by means of free linkers in solution \cite{rogers-manoharan,mcginley2013assembling}.}
%
 Halverson {\em et al.\ }\cite{Halverson2016} also  proposed the use of three-linker DNA complexes to program a cooperative behavior between functionalized particles, which could allow to control the sequence of binding events in the self-assembly.\\
{  Recently, Angioletti-Uberti {\em et al.\ }\cite{stefano-jcp}  proposed an analytical expression for the free energy of systems featuring one-to-one ligand-receptor interactions that overcame some limitations of earlier approaches \cite{melting-theory1,angioletti2016theory}}. In this Communication we provide a more general framework to calculate the free energy of systems including multimeric complexes featuring an arbitrary number of ligand/receptors  (see Fig.~\ref{Fig1}).
 We consider ``particles", e.g. biological cells or artificial colloidal units, functionalized by surface ligands/receptors (``linkers" or ``molecules''). We assume that linkers can freely diffuse on the surface of the particles. An extension to immobile linkers can be derived following Ref.\cite{patrick-jcp}. 
Bonds can either involve linkers tethered to the same particle or to different particles. Excluded volume interactions between the molecules are neglected. Our results are exact in the limit of many linkers per particle \cite{stefano-prl,xu2016simple}.
{ We envisage applications of our theory to the association of more complex molecules like DNA tiles\cite{dannenberg2015modelling,jacobsself,fern2016energy} or virial caspids \cite{dykeman2014solving}.}
\\
{ In Sec.~\ref{Sec:1} we derive our theory { while}
%{\cb In Sec.~\ref{Sec:1} we derive the general expression for the free energy of an ensemble of linkers forming multimeric complexes.}
in Sec.~\ref{Sec:2} we test it }on the system introduced in Ref.~\cite{Halverson2016},  calculating the interaction free energy between particles and quantitatively justifying the postulated cooperative behavior.
In Sec.~\ref{Sec:3} we examine the suspensions of DNA-functionalized vesicles of Ref.~\cite{Parolini2016}, discussing the thermodynamic ground state in relation to the kinetic behaviour characterized in the original publication.

\section{Free energy calculation}\label{Sec:1}

We consider $c$ families of different  linkers, each with a number of units $N_i$ ($i=1,\cdots ,c$).
The linkers can reversibly associate into complexes of $m$  units.  
For clarity we only consider complexes that never feature more than a single linker of each family ($1\leq m \leq  c$,  see Fig.~\ref{Fig1} where $c=3$). 
In Sec.\ S1 of the  supplemental material (SM)\footnote{See supplemental material at [URL will be inserted by AIP] for 
a more general derivation of Eq.\ \ref{eq:main} and for additional details of the examples studied in Sec.\ \ref{Sec:2} and Sec.\ \ref{Sec:3}. } 
we show that relaxing this assumption does not change our main result (Eq.\ \ref{eq:main}).
 The state of the system is described by the number $n_{i_1,i_2,\cdots i_m}$ of all the possible complexes made by $m$ linkers of type $i_1$, $i_2$, $\cdots$ $i_m$, with $i_1<i_2<\cdots i_m$ and $2\leq m \leq c$.
 \\
{ We start by deriving} an expression for the partition function { $Z$}
of the system as the { weighed} sum over all the possible realizations of  $n_{i_1,i_2,\cdots i_m}$.
% {\cb , then we derive the free energy using a saddle-point approximation.
%We evaluate the partition function recursively. } 
{ First} we calculate the contribution of two-linker complexes  $\{ n_{i_1,i_2} \}$ { to $Z$}, then we deplete the total number of linkers of each family $N_i$ by the 
 number of those involved in two-linker complexes and calculate the contribution from complexes with three molecules $\{n_{i_1,i_2,i_3}\}$. This procedure is repeated recursively.
 When calculating the contribution of  complexes with $m+1$ linkers,
 $N_i$  has been reduced to $N^{(m)}_{i}$ that is given by 
 \begin{eqnarray}
 N^{(m)}_i = N_i - \sum_{\ell=2}^{m} n^{(\ell)}_i
& \quad &
n^{(m)}_i = \sum_{i_2 < \cdots <i_m} n_{ \tau [i , i_2 , \cdots i_m]} 
\label{eq:N-depleted}
\end{eqnarray}
where $n^{(\ell)}_i$ is the total number of linkers of type $i$ involved in complexes of size $\ell$, and 
$\tau$ is the operator that orders $m$ indices. { $N^{(c)}_i$  is the number of linkers of type $i$ that are free, and will be 
also indicated by $n_i$ below. }
 The partition function is then expressed as 
 \begin{eqnarray}
Z= \sum_{  \{ n_{i_1,i_2} \} } Z^{(2)}  \sum_{  \{ n_{i_1,i_2,i_3} \} }  Z^{(3)}  \cdots   
\sum_{  \{ n_{i_1,i_2,\cdots i_c} \} } Z^{(c)}  ,
\label{zeta}
\end{eqnarray} 
where the curly brackets indicate the ensemble of all the complexes formed by a given number of linkers.
Note that in Eq.~\ref{zeta} $Z^{(\ell)}$ is a function of $\{ N^{(\ell-1)}_i\}_{1\leq i \leq c}$ 
 and, as a consequence of Eq.~\ref{eq:N-depleted}, of the number of complexes with $m\leq \ell$.
\\
Defining $\Delta G_{i_1, \cdots i_m}$ as the free energy associated to the formation of a 
$i_1 \cdots i_m$  complex \cite{dirks2007thermodynamic},
we can define the { contribution to the partition function from} all the complexes of size $m$ as
\begin{eqnarray}
Z^{(m)} &=&  \Omega^{(m)} \Big( \{ N^{(m-1)}_i \} ; \{ n_{i_1 , i_2 ,\cdots ,i_m} \}\Big)
 \label{eq:Zm} \\
&& \qquad \exp\Big[ 
 -\sum_{i_1<i_2 \cdots <i_m} n_{i_1,i_2 \cdots , i_m}  \beta\Delta G_{i_1 ,i_2 \cdots ,i_m}
 \Big],
\nonumber
\end{eqnarray}
where 
$\beta=1/(k_\mathrm{B} T)$ and
$\Omega^{(m)}$ accounts for the combinatorial factors. 
{ The latter can be written as}
%{\cb The latter are given by the following equation}
\begin{eqnarray}
\Omega^{(m)} =  \prod_{i=1}^c {N^{(m-1)}_i! \over N^{(m)}_i! } \prod_{i_1<i_2<\cdots <i_m} {1\over n_{i_1, i_2, \cdots, i_m} !},
\label{eq:Omega}
\end{eqnarray}
where the first product counts the number of ways  one can choose the molecules { belonging to} the complexes
$\{ n_{i_1, i_2, \cdots, i_m} \}$ starting from 
$\{ N_i^{(m-1)}\}$ free linkers, while  the second term accounts for the number of independent ways to build such a set of complexes.  Using Eq.~\ref{eq:Omega} and Eq.~\ref{eq:Zm} into Eq.~\ref{zeta}, we can calculate the partition function and the free energy $F$ of the system
\begin{eqnarray}
Z &=& e^{- \beta F} = \sum_{ \{  \{ n_{i_1, i_2, \cdots, i_\ell} \} \} }  \exp[-   \beta {\cal A}(\{  \{ n_{i_1, i_2, \cdots, i_\ell} \} \} ) ]
\label{zeta2}
 \\
 &=& \sum_{ \{  \{ n_{i_1, i_2, \cdots, i_\ell} \} \} }  \prod_{i=1}^c {N_i! \over n_i! } \prod_{m=2}^c \prod_{i_1<i_2<\cdots <i_m}{1\over n_{i_1, i_2, \cdots, i_m} !}
\nonumber
\\
&&\quad \quad \exp\Big[ 
 -\sum_{i_1<i_2 \cdots <i_m} n_{i_1,i_2 \cdots , i_m} \beta \Delta G_{i_1 ,i_2 \cdots ,i_m}
 \Big],
 \nonumber
\end{eqnarray} 
where the double curly brackets $\{\{\dots\}\}$ indicate the ensemble of complexes of arbitrary size, and ${\cal A}$ is a functional introduced for later convenience. 
\\
In the limit of large $N_i$ we can simplify Eq.~\ref{zeta2} using a saddle point approximation. In particular the stationary point of $\cal A$, given by $\delta {\cal A} / \delta n_{i_1,\cdots i_m}=0$, identifies the average number of complexes $\on_{i_1 i_2 \cdots i_m} = \langle n_{i_1 i_2 \cdots i_m} \rangle$. The stationary conditions  for the functional ${\cal A}$ as defined by Eq.~\ref{zeta2}  become 
%{\cb {\em if we need space this equation may be inline and we can refer to it as chemical equilibrium balance}}
\begin{eqnarray}
\on_{i_1 i_2 \cdots i_m} &=& \on_{i_1} \on_{i_2} \cdots \on_{i_m} \exp[-   \beta \Delta G_{i_1, \cdots i_m} ] \, \, \, .
\label{chemeq}
\end{eqnarray}
Note that Eq.\ \ref{chemeq} are the { relations} for chemical equilibrium expressed in terms of the total 
number of molecules. When considering tethered linkers (Fig.\ 1) the complexation free energy 
$\Delta G_{i_1, \cdots i_m}$ \cite{dirks2007thermodynamic} also includes 
rotational and translational entropic costs controlled by the length 
of the spacers and by the size of the particles\cite{parolini2014thermal}.
\\
Using Eq.~\ref{chemeq} into  Eq.~\ref{zeta2} to express $\Delta G_{i_1, \cdots i_m}$ as a function of { equilibrium number of complexes}, we can evaluate the free energy of the system as 
$F={\cal A}(\{  \{ \on_{i_1, i_2, \cdots, i_\ell} \} \} )$. 
{  By considering only the dominant term in the second line of Eq.\ \ref{zeta2}, and using Stirling's approximation we find }
%\begin{widetext}
\begin{eqnarray}
 \beta F &=& \sum_{i=1}^c \on_i \log \on_i  - N_i \log N_i - \on_i + N_i 
 \nonumber \\
&&  + \sum_{m= 2}^c \sum_{i_1<i_2 \cdots <i_m} \on_{i_1, i_2, \cdots, i_m}
\left( \log \on_{i_1} + \cdots \log \on_{i_m} -1\right)
\nonumber \\
 &=& { \sum_{i=1}^c N_i \log {\on_i \over N_i} - \on_i + N_i   - \sum_{m=2}^c \sum_{i_1< i_2\cdots <i_m} \on_{i_1, i_2 \cdots, i_m} }
 \nonumber \\
&=& \sum_{i=1}^c N_i \log { {\on_i \over N_i} } + \sum_{i=1}^c \sum_{m= 2}^c \sum_{i_2<i_3 \cdots <i_m} \on_{\tau[i,i_2, i_3, \cdots, i_m]} 
\nonumber \\
&& - \sum_{m =2}^c \sum_{i_1<i_2 \cdots <i_m} \on_{i_1, i_2, \cdots, i_m} \, ,
\nonumber 
\end{eqnarray}
{ where Eq.\ \ref{eq:N-depleted} has been used in the second equality to factorize the terms $\log \on_i$, and in the third equality to express $N_i - \on_i$ in terms of higher order complexes.  Finally we obtain the main result of this work}
\begin{eqnarray}
\beta F= \sum_{i=1}^c N_i \log {\on_i \over N_i }+  \sum_{m =2}^c (m-1) \sum_{i_1<i_2 \cdots <i_m} \on_{i_1, i_2, \cdots, i_m}.
\nonumber \\
\label{eq:main}
\end{eqnarray}
%\end{widetext}
%{\em we can delete the equation below or re-write 1 if necessary} }
%\begin{eqnarray}
%N_i = \on_i + \sum_{m= 2}^c \sum_{i_2<i_3 \cdots <i_m} \on_{\tau[i, i_2, \cdots, i_m]}
%\end{eqnarray}
Being written in term of the equilibrium number of complexes, { Eq.\ \ref{eq:main}}
 cannot be used to sketch free-energy landscapes \cite{jacobsself,fern2016energy}, however
it { is applicable to calculate} effective interactions between functionalized objects as 
{ demonstrated} in the next two sections.
 %{\cm If only dimers are allowed},
%
Note that in order to guarantee the extensivity of the functional ${\cal A}$, $\Delta G_{i_1, ... i_m}-(m-1)\log N$ (with $N_i\sim N$) should be kept fixed when taking the $N\to \infty$ limit (see also Eq.\ \ref{chemeq}). 
For the case of one-to-one interactions  ($m=2$), Eq.~\ref{eq:main} reduces to the result of Ref.\ \cite{stefano-jcp}.
%{\cb For fixed $\Delta G_{i_1, ... i_m}$ the fraction of associated complexes increases as a function of $N$ for combinatorial reasons \cite{parolini2014thermal}.}

\begin{figure}
\includegraphics[width=8.75cm]{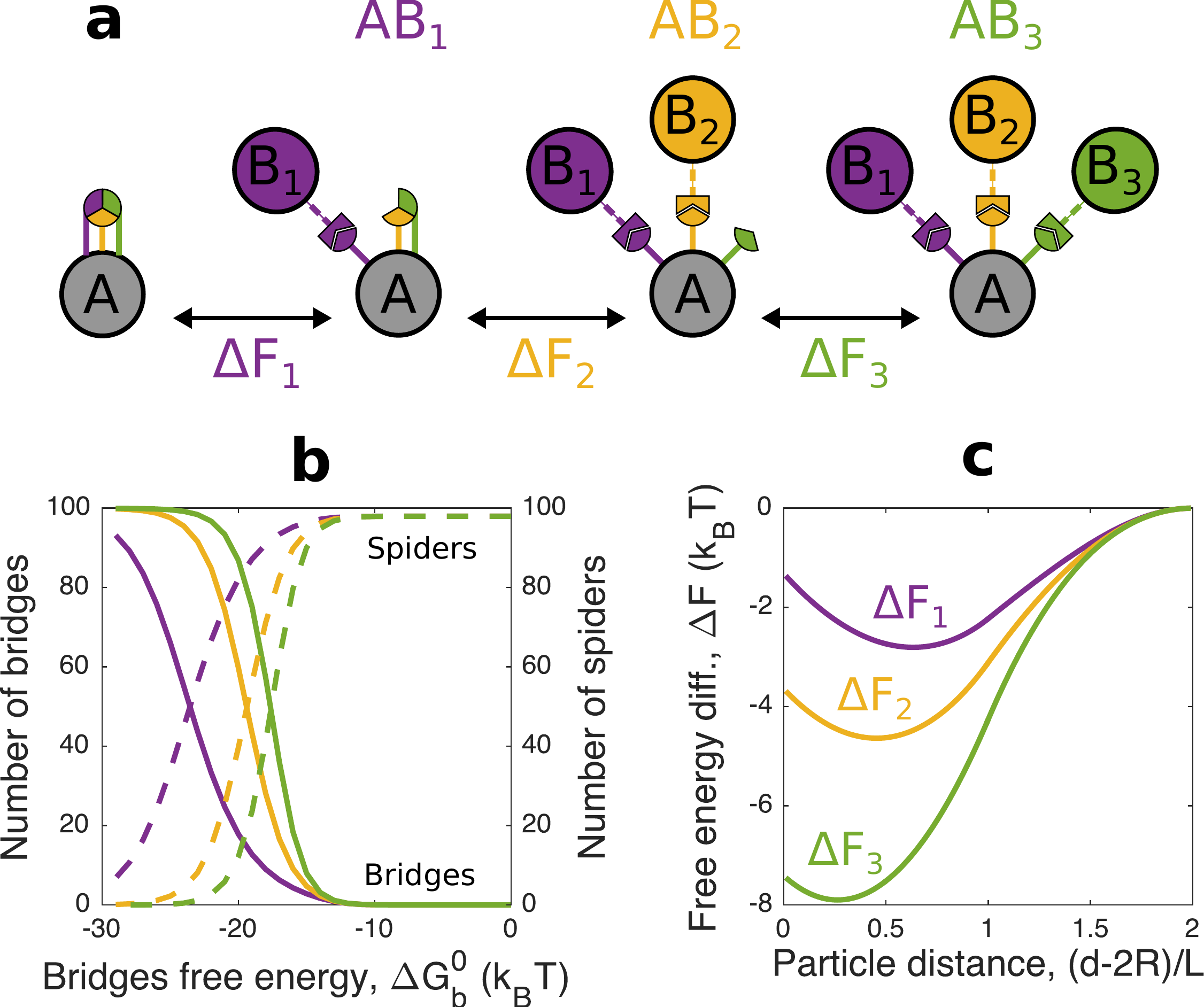} 
\caption{ 
Cooperative binding scheme of Ref.~\cite{Halverson2016}. \textbf{a}, For isolated $A$ particles most linkers form spiders. For $B_1$ to bind, stable spiders need to break to form bridges. Loops are left on $A$. The binding of $B_2$ is more favourable, as loops are less stable than spiders. Binding the third particle is most favorable, as no intra-particle complexes need to open on $A$. \textbf{b}, Number of spiders and bridges in $AB_n$ complexes, color-coded following panel \textbf{a}. \textbf{c}, Free energy difference between the complexes shown in panel \textbf{a} as a function of inter-particle distance for $\Delta G^0_b=-14 k_B T$.
%%{\cb
%%Cooperative binding scheme considered in Ref.~\cite{Halverson2016}. \textbf{a}, In an isolated $A$ particle most linkers form three-strand complexes (spiders). For $B_1$ to bind stable spiders need to break to form bridges. Loops are left on $A$. The binding of $B_2$ is more favourable, as loops are less stable than spiders. Binding the third particles is most favorable, as no intra-particle complexes need to open on $A$. \textbf{b}, Number of spiders and bridges in $AB_n$ complexes, color-coded following panel \textbf{a}. \textbf{c}, Free energy difference between the complexes shown in panel \textbf{a} as a function of inter-particle distance for $\Delta G^0_b=-14 k_B T$.
% %}
 \label{fig:spider}
 }
 \end{figure}

\section{ Binding cooperativity in DNA-functionalized particles }\label{Sec:2}
As a first example we examine { the} cooperative self-assembly scheme recently proposed by Halverson and Tkachenko\cite{Halverson2016},  
based on the possibility of forming three-linker complexes, dubbed spiders. 
%{\cm (see Fig.\ \ref{fig:spider}{\bf a})}. 
%{\cb The  system's design is sketched in Fig.\ \ref{fig:spider}{\bf a}. }
%
{ As shown in Fig.~2\textbf{a}, we} consider particles of type $A$  functionalized by 3$N$ mobile DNA linkers equally distributed among three families, 
 each carrying different single-stranded DNA sequences or \emph{sticky-ends}, labelled as $\alpha_i$, $i=1,\, 2, \, 3$.
Such sticky-ends can hybridize to form three different families of intra-particle \emph{loops} ($\ell_i$), involving two out of three types of linkers, or spiders ($s$), involving all three types (see Fig.~\ref{fig:spider}\textbf{a}). 
%We refer to SI Fig.\ ??  and to Ref.\ [] for a possible architecture of the $\alpha_i$ domains. 
%
We then consider three types of particles $B_i$, $i=1,2,3$, each functionalized by $N$ identical linkers carrying 
a sticky-end sequence $\overline \alpha_i$ complementary to $\alpha_i$. Linkers on particles $B_i$ can form inter-particle \emph{bridges} $b_i$, with particles $A$.
In the following we consider linkers constituted by double stranded DNA spacers of length $L=10\,$nm and point-like sticky ends\cite{stefano-prl}, rigid particles of radius $R=10\,L$\cite{stefano-prl}, and $N=100$.  See SM Sec.\ S2 and Ref.~\cite{stefano-prl} for details.
%The  condition $L\ll R$ simplifies the calculation of the configurational costs affecting the hybridization free energy of tethered DNA strands \cite{stefano-prl}. 
%For further details we refer to the SI Sec.\ S2 and to Ref.~\cite{stefano-prl}.\\
Below we calculate the free energy $F(AB_n)$ of clusters made by a single $A$  particle and a variable number $n$ of $B$ particles taken as in Fig.\ \ref{fig:spider}.  We demonstrate a cooperative effect by which  the free energy gain from binding the  $n$-th $B$ particle $\Delta F_n = F(AB_n)-F(AB_{n-1})$ is higher than the gain from binding the $(n-1)$-th one, for $n=2,3$. This is due to the necessity of breaking %(eventually stable) 
spider and loop complexes formed on the $A$ particle for the 1st or the 2nd $B$ particles to bind. 
Our theory allows to calculate the free energy gain for binding the 1st, the 2nd and the 3rd $B$ particles, chosen as a model parameters in Ref.~\cite{Halverson2016}. 
%
%{\cb {\em this sentence can also be deleted} }
%We also note that a negative cooperative behaviour, where the free energy gap of attaching later particles is negative,
%has been proposed as a means of controlling valence in a system featuring mobile DNA linkers and inert repulsive stands\cite{stefano-prl}.\\
The number of complexes at equilibrium are given by \cite{stefano-prl}
\begin{eqnarray}
\on_{\ell_k} &=& \on_{\alpha_i} \on_{\alpha_j} \Big[ {e^{-\beta \Delta G^0_\ell} \over \rho_\ominus v_A} \Big]
\quad (k\neq i, j\,\, \mathrm{AND} \, i\neq j)
\label{loops}
\\
\on_s &=& \on_{\alpha_1} \on_{\alpha_2} \on_{\alpha_3} \Big[ {e^{-\beta \Delta G^0_s} \over (\rho_\ominus v_A)^2} \Big]
\label{spider}
\\
\on_{b_i} &=& \on_{\alpha_i} \on_{\bar{\alpha}_i}   \Big[ {\epsilon_i v_{AB} e^{-\beta \Delta G^0_b} \over \rho_\ominus v_A v_B } \Big]
\label{bridges}
\end{eqnarray}
where $\Delta G^0_{\ell}$, $\Delta G^0_{s}$ and $\Delta G^0_{b}$ are the hybridization free energies
of the sticky-ends associated to loop, spider, and bridge formation respectively,
 $\rho_\ominus$=1M is the standard concentration, and $v_{A/B/AB}$ are volume factors reported in SM Sec.~S2 that quantify the configurational entropic costs of binding  mobile tethers (Refs.\ \cite{parolini2014thermal,stefano-prl} and SM Sec.\ I). 
{ Note that by} using the generic notation of Sec.\ \ref{Sec:1} we would have $\on_{\ell_k}  \equiv   \on_{\alpha_i \alpha_j} $, $\on_s 
 \equiv  \on_{\alpha_1 \alpha_2 \alpha_3}$, and $\on_{b_i}  \equiv  \on_{\alpha_i \bar{\alpha}_i} $.
Different types of loops and bridges are assumed to have the same hybridization free energy.
In Eq.~\ref{bridges} $\epsilon_i = 0,1$ specifies if $B_i$ is bound or not to $A$. In particular $n=\sum_i \epsilon_i$ indicates the 
valence of particle $A$. 
Equations~\ref{loops}, \ref{spider}, and \ref{bridges} are then closed by the conditions $N = \on_{\alpha_i} + \on_{\ell_j} +\on_{\ell_k}  + \on_s + \on_{b_i}$ and $N= \on_{\overline{\alpha}_i} +  \on_{b_i}$.

First we consider an isolated $A$ particle and calculate the number of loop and spider complexes as a function 
of $\Delta G^0_\ell$, choosing $\Delta G^0_s=3 \Delta G^0_\ell$. As shown in Fig.~S1 of the SM,
when $\Delta G^0_\ell=-10\, k_B T$
only spiders are present { on $A$}. We fix $\Delta G^0_\ell$ to this value as a reasonable guess to maximize the 
cooperative behaviour.
We then consider particle clusters $A B_n$ (with $n=0,1,2,3$), with distances between the { centers of} $A$ and $B$ particles equal to $d=2R+L$, and calculate the number of bridges $n_{b_i}$ as a function of  $\Delta G^0_b$ (see Fig.\ \ref{fig:spider}{\bf b}). 
We find that bridges form at higher values of $\Delta G^0_b$ when $n$ is higher. 
Finally we use Eq.~\ref{eq:main} (contextualized to this system in SM Eq.~S14) to calculate the free energy of the system including the repulsive part of the potential calculated accounting for the entropic compression of the DNA strands between the particles (see SM Eq.~S2). 
We consider clusters in which { all of the $B$ particles are at the same distance $d$ from the $A$-particle,} and for which $B$-particles do not interact with each other (see Fig.~\ref{fig:spider}\textbf{a}).  Figure~\ref{fig:spider}\textbf{c} shows the free-energy change associated to the binding of a single $B$ particle to a cluster as a function of $d$. As expected, the free-energy gain obtained when adding the second $B$ particle is higher than that obtained by binding the first, and the gain achieved upon adding the third particle is significantly higher than both the former.\\
We note that kinetic bottlenecks associated to the opening of the stable spider and loop { complexes} are likely  { to slow down self-assembly}. Incidentally, strand-displacement strategies\cite{zhang2009control} similar to those discussed in the next section and in Ref.\cite{Parolini2016} can speed up relaxation.

\begin{figure}
\includegraphics[width=8.75cm]{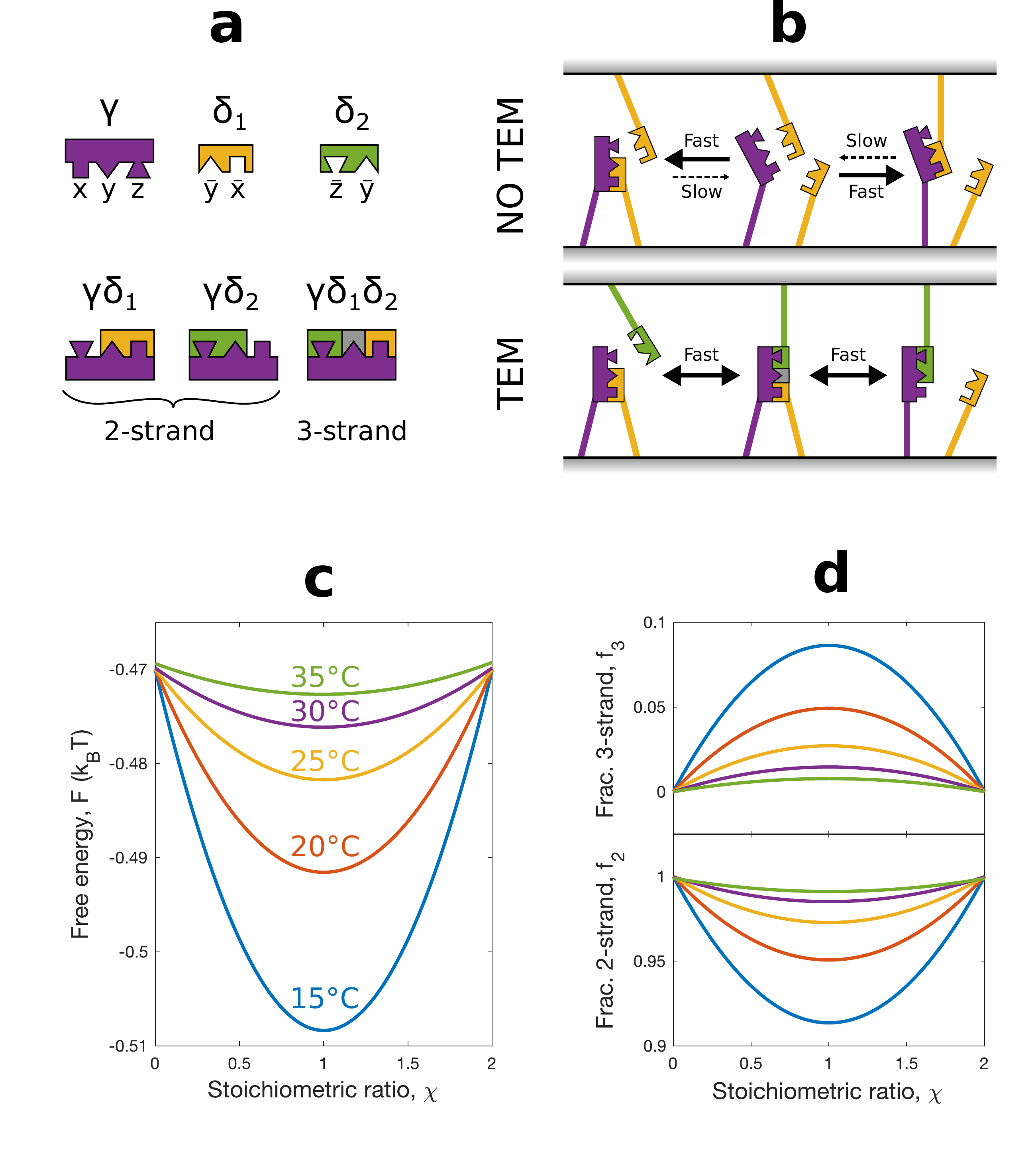} 
\caption{ \label{fig:toeholding} Toehold-exchange mechanism of Ref~\cite{Parolini2016}. \textbf{a}, We consider three  sticky ends $\gamma$, $\delta_1$ and $\delta_2$ that can form { two}- or { three}-strand complexes. \textbf{b}, If only one between $\delta_1$ and $\delta_2$ are present ($\chi=0,2$), the formation of an inter-particle bridge requires the thermal breakup of an intra-particle bridge, making aggregation slow (Top). If both $\delta_1$ and $\delta_2$ are present, TEM catalyzes loop-to-bridge swap. \textbf{c}, Despite the difference in kinetic behavior, the free energy of the system { (\emph{per} $\gamma$ strand)} as calculated within our framework depends on $\chi$ only weakly regardless of temperature. \textbf{d}, Fraction of { two-strand ($f_2$)} and { three-strand ($f_3$)} complexes as calculated by our theory at different temperatures.
}
\end{figure}

\section{Interaction free energy in the presence of Toehold-Exchange-Mechanism}\label{Sec:3}

As a second example we examine the system studied experimentally in Ref.~\cite{Parolini2016}.  
 Let us consider a suspension of identical micron-size lipid vesicles, functionalized by three families of mobile DNA linkers with sticky ends here labelled as $\gamma$, $\delta_1$, and $\delta_2$. As shown in Fig.~\ref{fig:toeholding}\textbf{a}, sticky end $\gamma$ is made of three domains of equal length, $x$, $y$, and $z$. Sticky ends $\delta_1$ have two domains $\bar{x}$ and $\bar{y}$, complementary to $x$ and $y$, whereas $\delta_2$ features domains $\bar{y}$ and $\bar{z}$. Linker $\gamma$ can bind to $\delta_1$ and $\delta_2$, with comparable hybridization free energy. A three-linker $\gamma \delta_1 \delta_2$ complex is also possible, where $\delta_1$ and $\delta_2$ bind to domains $x$ and $z$ respectively, and compete to occupy domain $y$. $\delta_1$ does not bind to $\delta_2$. Two- and three-linker complexes can form either among linkers tethered to the same vesicles (loop-like) or between different vesicles (bridge-like).
At sufficiently high temperature all of the linkers are unbound. If the suspension is quenched to low temperature, the formation of intra-vesicle loop-like { complexes} is kinetically favored over the formation of bridges, effectively sequestrating all of the available $\gamma$ linkers. The aggregation of the liposomes, mediated by the formation of inter-vesicle bridges, is therefore limited by the opening of the intra-vesicle loops, which seldom occurs at low temperatures (Fig.~\ref{fig:toeholding}\textbf{b}, Top). Through a \emph{Toehold-Exchange Mechanism} (TEM)\cite{zhang2009control}, the formation of three-strand { complexes} mediates the swap between stable loops and stable bridges without the need for thermal denaturation. In particular, the toehold domain $z$ ($x$)  causes a free $\delta_2$ ($\delta_1$) linker to transiently bind to an existing $\gamma\delta_1$ ($\gamma\delta_2$) bond, facilitating the reaction $\gamma \delta_1 + \delta_2 \leftrightharpoons \delta_1 + \gamma \delta_2$ (Fig.~\ref{fig:toeholding}\textbf{b}, Bottom).
 We indicate with $3N$ the total number of linkers per vesicle, $N$ of which are of type $\gamma$, $N\chi$ of type $\delta_1$, and $N(2-\chi)$ of type $\delta_2$. The parameter $\chi \in [0,2]$ controls the stoichiometric ratio between $\delta_1$ and $\delta_2$ and thereby the effectiveness of the TEM process. For $\chi=0$ or $2$ three-strand 
 { complexes} are not possible and the bridge formation and aggregation kinetics are dominated by the slow opening of formed loops. For $\chi=1$, TEM is most effective and aggregation kinetics is found to speed up by more than one orderer of magnitude at $T=15^\circ$C\cite{Parolini2016}. \\
 We { use} our framework to calculate the { free energy of the system}, and demonstrate that, despite the large effect on aggregation kinetics, changing $\chi$ has little consequences on the thermodynamic ground state of the system. The DNA tethers are again modelled as freely pivoting rigid rods of length $L=10$ nm, with freely diffusing tethering points and point-like sticky ends. For simplicity we model two interacting vesicles as flat planes of area $A=0.5\mu$m$^2$ kept at a distance of { $h=1.4 L$} from each other. We chose $N=360$. Hybridization free-energies between the sticky ends are taken from Ref.\cite{Parolini2016}. \\
Explicit expression for the equilibrium distributions of all the possible complexes are shown in the SM Eqs.~S15-20. In the SM (Eqs. S21, S22) we provide the expression for the interaction free energy between two vesicles { (\emph{per} $\gamma$ strand)}, shown as a function of $\chi$ and $T$ in Fig.~\ref{fig:toeholding}\textbf{c}. We observe that regardless of temperature, the free energy decreases { by less than} 10\%
when going from $\chi=0, 2$ to $\chi=1$, supporting the claim that { with} the architecture proposed in Ref.\cite{Parolini2016} aggregation kinetics can be substantially changed with little consequences on the thermodynamic ground state. The weak dependence of the overall free energy on $\chi$ is a direct consequence of the small { number of} three-strand { complexes}, always involving less { than $10\%$} of all $\gamma$ linkers, as demonstrated in Fig.~\ref{fig:toeholding}\textbf{d}.

\section{Conclusions}

We provide an analytical expression for the free energy of systems of ligand/receptors that can form complexes featuring an arbitrary number of molecules. 
Our framework can be applied to biologically relevant situations, where cell-surface receptors form { trimers} or oligomers, or to suspensions of colloidal particles functionalized by synthetic DNA ligands: an increasingly popular strategy to achieve controlled self-assembly of complex soft materials.
To exemplify the versatility of our approach, we re-examine the artificial systems recently proposed by Halverson {\em et al.\ }\cite{Halverson2016} and Parolini {\em et al.\ }\cite{Parolini2016}, both featuring DNA-functionalized Brownian particles interacting trough the formation of three-linker complexes. For the former, we are able to quantify the cooperative effects in the interaction free energy between the particles, taken as model parameters in the original publication. For the system of Parolini {\em et al.\ } we study the interaction free energy between vesicles with different linker stoichiometry. Our theory  demonstrates that despite the substantial effect on aggregation kinetics observed experimentally, coating stoichiometry has a comparatively small effect of the thermodynamic ground state of the suspension.\\

{\bf Acknowledgements. } LDM and LP acknowledge support from the EPSRC Programme Grant CAPITALS number EP/J017566/1. LDM acknowledges support from the Oppenheimer Fund and Emmanuel College Cambridge. SB and BMM are supported by the Universit\'e Libre de Bruxelles (ULB). The python programs used to calculate the results presented here are available at \url{http://dx.doi.org/10.5281/zenodo.47204}.

\end{document}

\title{SUPPLEMENTAL MATERIAL: Free energy of ligand-receptor systems forming multimeric complexes}
 \author{Lorenzo Di Michele}
\affiliation{Biological and Soft Systems, Cavendish Laboratory, University of Cambridge, JJ Thomson Avenue, Cambridge, CB3 0HE, United Kingdom}
\author{Stephan J.\ Bachmann}
\affiliation{Interdisciplinary Center for Nonlinear Phenomena and Complex Systems \&
Service de Physique des Syst\`{e}mes Complexes et M\'{e}canique Statistique, Universit\'{e} libre de Bruxelles (ULB), Campus Plaine, CP 231, Blvd du Triomphe, B-1050 Brussels, Belgium.}
\author{Lucia Parolini}
\affiliation{Biological and Soft Systems, Cavendish Laboratory, University of Cambridge, JJ Thomson Avenue, Cambridge, CB3 0HE, United Kingdom}
\author{Bortolo M. Mognetti}
\email{bmognett@ulb.ac.be}
\affiliation{Interdisciplinary Center for Nonlinear Phenomena and Complex Systems \&
Service de Physique des Syst\`{e}mes Complexes et M\'{e}canique Statistique, Universit\'{e} libre de Bruxelles (ULB), Campus Plaine, CP 231, Blvd du Triomphe, B-1050 Brussels, Belgium.}

\maketitle

\section{Reactions between binders of the same type}
 
 In this section we relax the hypothesis by which each complex cannot feature more than a single linker of a given type and  re--derive Eq.\ 7 of the main text. When specifying a given complex made by $i_1$, $i_2$, $\cdots$, $i_m$ linkers  ($X=\{i_1,\cdots i_m \}$), we now only assume that $i_1\leq i_2 \cdots \leq i_m$.
 The number of { linkers} of type $i$ entering the complex $X=\{i_1,\cdots \, ,i_m\}$
  is defined as 
 \begin{eqnarray}
g_i(X) = g_i(\{i_1,\cdots i_m \}) = \sum_{\alpha=1}^m \delta_{i,i_\alpha} \, \, .
 \end{eqnarray}
 In the following with $\{\{X\}\}$ we refer to the ensemble of all possible complexes with at least two linkers while with 
 $\{ X\}_m$ we  refer to the ensemble of complexes made by $m$  linkers.
 Using these definitions it is not difficult to show that the partition function of the system (Eq.\ 5, main text) becomes
 \begin{eqnarray}
Z= \sum_{ \{ \{ n_X \} \}} \prod_{i=1}^c {N_i! \over n_i !} \prod_{m\geq 2} \prod_{\{ X \}_m}{1\over n_X ! \left[ \prod_{j=1}^c g_j(X)! \right]^{n_X}} \exp[-n_X \beta \Delta G_X] \, \, .
 \label{eq:Z}
 \end{eqnarray}
{ Note} that in Eq.\ \ref{eq:Z} we { do} not distinguish { between} the $g_i$ monomers { in} a given complex. 
This is not justified in systems featuring structured complexes where identical monomers can bind with different free energy depending on the site they occupy within the complex. This scenario may occur in nucleic acid complexes featuring several strands\cite{dirks2007thermodynamic}, isomeric clusters in gelation theory \cite{gordon1964non}, or polymerization \cite{andrieux2008nonequilibrium}.
{ For the purpose of the present work this scenario would not change the final result 
in view of the fact that we derive an expression for the free energy of the system in which the binding free energy 
of the single complex $\Delta G_X$ 
is expressed in terms of equilibrium densities { (see Eq.\ \ref{eq:saddle})} and of the fact that different combinatorial factors of 
the complexes would simply re-define $\Delta G_X$ in Eq.\ \ref{eq:Z}.}

Using Eq.\ \ref{eq:Z} we can calculate the functional ${\cal A}$ defined in Eq.\ 5  of the main text
 \begin{eqnarray}
 \beta {\cal A}(\{\{ n_X \}\}) &=& \sum_{i=1}^c \left[ n_i \log n_i - n_i -N_i \log N_i + N_i \right]
 +\sum_{m\geq 2} \Bigg[
 \sum_{\{X\}_m} \Big[ 
 n_X \log n_X - n_X
 \nonumber \\
 && +n_X \log \Big( \prod_{j=1}^c g_j(X) ! \Big)
+ n_X \beta \Delta G_X
 \Big]
 \Bigg]
 \label{eq:functional}
 \end{eqnarray}
 where the number of free binders of type $i$ ($n_i$) is written as 
 \begin{eqnarray}
 n_i &=& N_i - \sum_{\{X\} } g_i(X) n_X \, .
 \label{eq:balance}
 \end{eqnarray}

 The stationary equations $\delta {\cal A} / \delta n_X = 0$ providing the equilibrium distribution $\on_X$ are $(\forall \, X)$
 \begin{eqnarray}
 \beta \Delta G_X + \log \Big( \prod_{i=1}^c g_i(X) ! \Big) + \log \on_X - \sum_{i=1}^c g_i(X) \log \on_i =0\, .
 \label{eq:saddle}
 \end{eqnarray}
 where we used Eq.\ \ref{eq:balance}. Notice in particular that Eqs.\ \ref{eq:saddle} can be rewritten into a standard equilibrium  balance
 \begin{eqnarray}
 {\on_X \over \prod_{i=1}^c  \on_i^{ g_i(X) }} &=& {e^{-\beta \Delta G_X} \over \prod_{i=1}^c g_i(X) !} \,. 
 \end{eqnarray}
 Using Eq.\ \ref{eq:saddle} multiplied by $\on_X$ in Eq.\  \ref{eq:functional} we obtain the free energy of the system as a function of equilibrium distributions $\on_X$
 \begin{eqnarray}
 \beta F &=& \sum_{i=1}^c \left[ \on_i \log \on_i - \on_i -N_i \log N_i + N_i \right] + \sum_{m\geq 2} \Bigg[
 \sum_{\{X\}_m} \Big[ \on_X \sum_{i=1}^c g_i(X) \log \on_i  - \on_X \Big] \Bigg]
 \nonumber \\
 &=& \sum_{i=1}^c \Bigg[\Big[\on_i +\sum_{\{X\}} \on_X g_i(X)\Big]\log \on_i -N_i \log N_i + [N_i-\on_i]\Bigg] -\sum_{\{ \{X \}\}} \on_X
 \nonumber \\
 &=& 
 \sum_{i=1}^cN_i \log {\on_i \over N_i} + \sum_i \sum_{\{X\} } g_i(X) \on_X  -\sum_{\{\{ X\} \}} \on_X
 \nonumber \\
 &=& \sum_{i=1}^cN_i \log {\on_i \over N_i} +\sum_{m\geq 2} \sum_{\{ \{X\} \}_m} (m-1) \on_X
 \label{eq:free}
 \end{eqnarray}
 where we have used multiple times Eq.\ \ref{eq:balance} and the fact that $\sum_i g_i(X)=m$ if $X\in \{ X\}_m$.  Note that Eq.\ \ref{eq:free} has the same functional form of  Eq. 7 of the main text.

\section{Binding cooperativity in DNA-functionalized particles}

We define by $\alpha (R_1,R_2,d)$ the volume of the intersection between two spheres of radius $R_1$ and $R_2$ with their center placed at distance equal to $d$. Defining $\sigma_+=R_1+R_2$ and $\sigma_-=|R_2-R_1|$ we have 
\begin{eqnarray}
\alpha(R_1,R_2,d) &=& {\pi \over 12 d} (\sigma_+-d)^2(d^2+2d\sigma_+-3\sigma_-^2)  \qquad \qquad \sigma_- <d< \sigma_+
\end{eqnarray}
Using the previous equation we can calculate the repulsive part of the potential due to entropic compression of the tethered DNA linkers \cite{stefano-prl}. In particular
we find 
\begin{eqnarray}
 \beta F_\mathrm{rep} &=& -3 N \log \left[ 1-  n {\alpha(R+L,R,d)\over \Omega_\infty} \right]
-N  n \log \left[ 1-  {\alpha(R+L,R,d)\over \Omega_\infty} \right]
\label{eq:free_rep}
\end{eqnarray}

\begin{figure}[ht!]
\includegraphics[width=10cm]{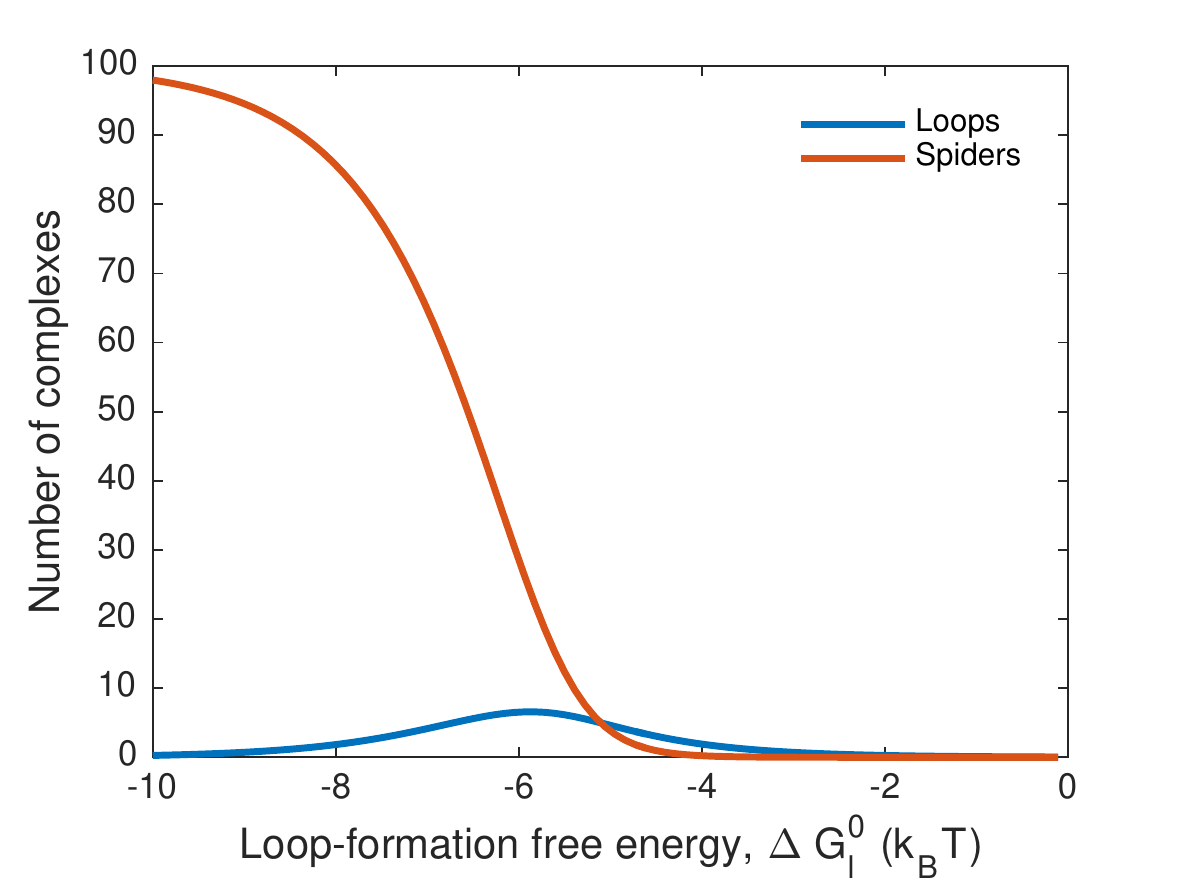} 
%\vspace{-0.5cm}
\caption{ Total number of loop and spider { complexes} as a function of $\Delta G^0_\ell$ for an isolated $A$ particle in which 
bridges cannot form. The free energy of the spider sticky-end { complex} is taken equal to $\Delta G^0_s=3 \Delta G^0_\ell$ 
as justified by the spider architecture suggested by Halverson and Tkachenko\cite{Halverson2016} formed by the hybridization of three complementary fragments of DNA, while a single hybridization directs the formation of loops. { Note} that such estimate neglects stacking terms and inert-tail effects that may be considerable.\cite{lorenzo-jacs}
{ Note} also that it is easy to foresee more { complex} sticky-end designs that would allow to tune $\Delta G^0_\ell$ and $\Delta G^0_s$ more independently. 
 \label{fig:loops_spiders}
 }
 \end{figure}

In Eqs.\ (9-11)  of the main text $v_A$, $v_B$, and $v_{AB}$ are the volume available to the sticky ends free to move on particle $A$, on particles $B_i$ (when close to particle  $A$), and when bridging  particle  $A$ with particles $B_i$ respectively. 
In particular we find
\begin{eqnarray}
v_A &=& \Omega_\infty -  n \alpha (R+L,R,d)
\nonumber \\
v_B &=& \Omega_\infty - \alpha (R+L,R,d)
\nonumber \\
v_{AB} &=& \alpha (R+L,R+L,d) - 2  \alpha (R+L,R,d)
\label{eq:volume}
\end{eqnarray}
where, as defined in the main text, $n$ is the number of particles $B_i$ attached to particle $A$, and $\Omega_\infty$ is the space availbale to the sticky ends on isolated particles
\begin{eqnarray}
\Omega_\infty= {4  \pi \over 3} \left[ (R+L)^3 - R^3 \right] \, .
\label{eq:Omegainf}
\end{eqnarray}
{Note} that Eqs.\ \ref{eq:free_rep}, \ref{eq:volume}, and \ref{eq:Omegainf} have been derived 
in the limit of $L\ll R$ and for double stranded DNA spacers modelled as rigid rods. Only when these assumptions hold the sticky ends are uniformly 
distributed in the layer between two spheres or radii $R$ and $R+\ell$ ~\cite{stefano-prl}. 
 For further geometrical assumptions we refer to the SI of Ref.~\cite{stefano-prl}.
If we define
\begin{eqnarray}
\Xi_\ell(d) = {e^{-\Delta G^0_\ell} \over \rho_\ominus v_A} & \qquad\qquad  \Xi_s (d) =  {e^{-\Delta G^0_s} \over (\rho_\ominus v_A)^2} \qquad\qquad & \Xi_b (d) ={v_{AB} e^{-\Delta G^0_b} \over \rho_\ominus v_A v_B }
\end{eqnarray}
Eqs. (8--10) of the main text  can then be rewritten as  (assuming $i\neq j$, $i\neq k$, and $j\neq k$)
\begin{eqnarray}
\on_{\alpha_i} &=& {N \over 
1+ (\on_{\alpha_j} + \on_{\alpha_k}) \Xi_\ell +  \on_{\alpha_j} \on_{\alpha_k} \Xi_s + \epsilon_i \on_{\overline{\alpha}_i} \Xi_b
}
\nonumber
\\
\on_{\overline{\alpha}_i} &=&  { N \over 1 + \epsilon_i \on_{\alpha_i} \Xi_b }
\label{eq:sc2}
\end{eqnarray}
We numerically solve Eqs.\ \ref{eq:sc2} and use Eqs.\ 9-11 of the main text to calculate the fraction of hybridized strands.
Results are given in Fig.\ \ref{fig:loops_spiders} and Fig.~2\textbf{b} of the main text.
 In particular Fig.\ \ref{fig:loops_spiders} reports the number of loops and spiders for an isolated 
$A$ particle ($n=0$) as given in Sec.\ II of the main text.

\begin{figure}[ht!]
\includegraphics[width=10.0cm]{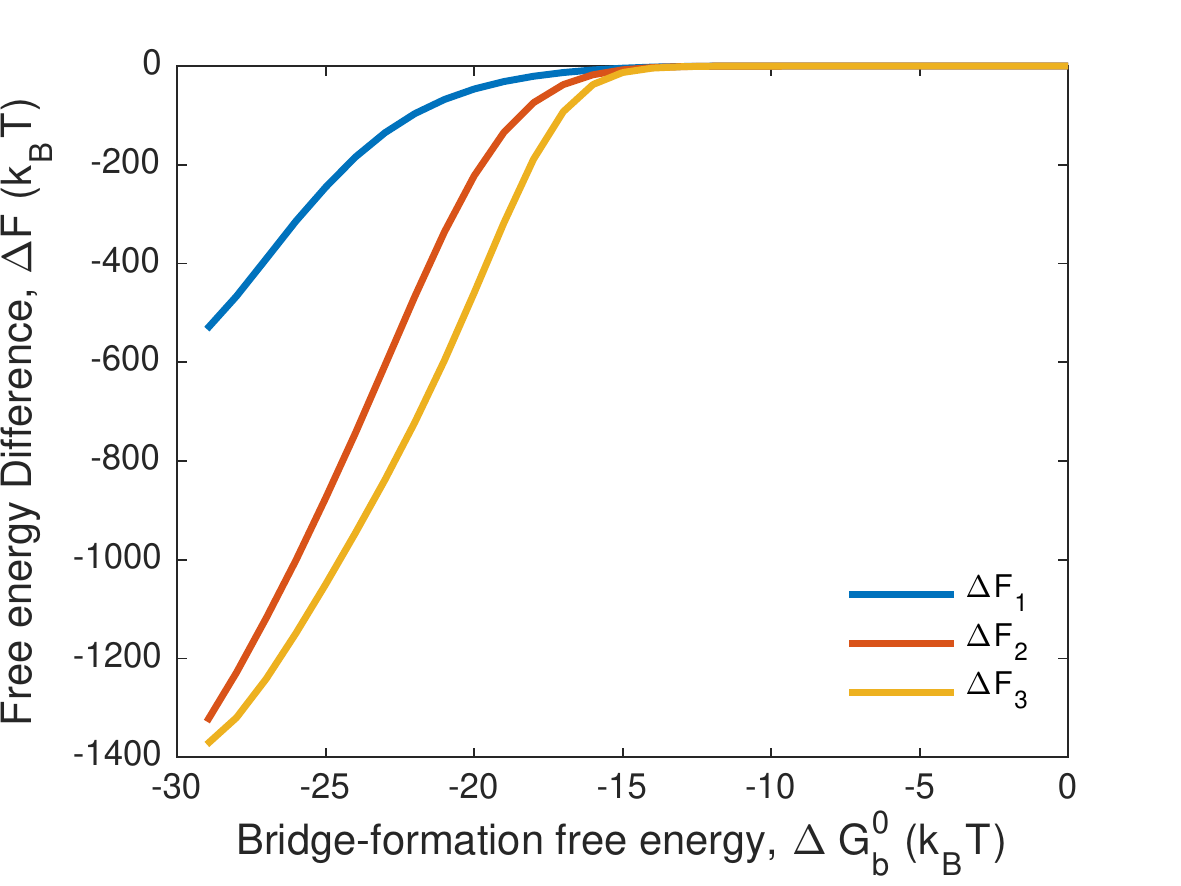}
%\vspace{-0.5cm}
\caption{
Free energy difference between the complexes shown in Fig. 2{\bf a} of the main text with { $d=2R+L$} as a function 
 of the hybridization free energy of the sticky ends responsible for the formation of bridges.  \label{fig:spiders_free} 
  }
 \end{figure}

  Appying Eq. 7 of the main text to this system we can then calculate the selective part of the interaction free energy 
\begin{eqnarray}
\beta F= N \sum_{i=1}^3 \left[ \log {\on_{\alpha_i}\over N} + \log {\on_{\overline{\alpha}_i}\over N} \right] +  \sum_{i=1}^3 \left[ 
\on_{\ell_i} + \on_{b_i} \right] + 2 \on_s.
\label{eq:free_att}
\end{eqnarray}
The overall interaction free energy is then calculated by adding up Eq.\ \ref{eq:free_att} to the steric repulsion described by  Eq.\ \ref{eq:free_rep}. The results are shown in Fig.~2\textbf{c} of the main text and Fig.~ \ref{fig:spiders_free}\\

 \section{Interaction free energy in the presence of Toehold-Exchange-Mechanism}

The toehold system introduced in Ref.\ \cite{parolini2016} and summarized in Sec.\ II of the main text features four types of two-strand complexes (see also Fig.\ 3 of Ref.\ \cite{parolini2016}): $\ell_{1/2}$ are loops due to the hybridization 
of $\delta_{1/2}$ with $\gamma$, while $b_{1/2}$ are bridges due to the hybridization 
of $\delta_{1/2}$ with $\gamma$.
The average number of two strand complexes is then given by
\begin{eqnarray}
\on_{\ell_1} &=&{ \on_{\delta_1}   \on_\gamma \over \rho_\ominus L A } \exp[-\beta \Delta G^0_{\gamma\delta_1}]
\nonumber \\
\on_{\ell_2}  &=& { \on_{\delta_2}   \on_\gamma  \over \rho_\ominus L A } \exp[-\beta \Delta G^0_{\gamma\delta_2}]
\nonumber \\
\on_{b_1} &=& { \on_{\delta_1}   \on_\gamma  \over \rho_\ominus L A } \Big( 2-{h\over L} \Big) \exp[-\beta \Delta G^0_{\gamma\delta_1}]
\nonumber \\
\on_{b_2}  &=&{ \on_{\delta_2}   \on_\gamma  \over \rho_\ominus L A } \Big( 2-{h\over L} \Big) \exp[-\beta \Delta G^0_{\gamma\delta_2}]
\end{eqnarray}
where $\Delta G^0$ are the hybridization free energies of the free sticky-ends in solution (we refer to the SI of Ref.\ \cite{parolini2016} for their value). { Following Ref. \cite{parolini2016} we take the inter-membrane distance as $h=1.4L$.} The bottom panel of Fig.\ 
3{\bf d} in the main text reports the amount of two-strand complexes {\em per} $\gamma$ strand
\begin{eqnarray}
{ f_2} = {\on_{b_1}+\on_{b_2} +\on_{\ell_1} +\on_{\ell_2} \over N }
\end{eqnarray}
For isolated vesicles the bridge { complexes} are not possible and we have 
\begin{eqnarray}
\on^0_{\ell_1} &=&{ \on^0_{\delta_1}   \on^0_{\gamma} \over \rho_\ominus L A } \exp[-\beta \Delta G^0_{\gamma\delta_1}]
\nonumber \\
\on^0_{\ell_2}  &=& { \on^0_{\delta_2}   \on^0_{\gamma}  \over \rho_\ominus L A } \exp[-\beta \Delta G^0_{\gamma\delta_2}]
\end{eqnarray}
We have four types of three-strand complexes, three bridging the two vesicles ($t_1$, $t_2$, and $t_B$) { and} the fourth ($t_3$) 
featuring a double loop structure (see Fig.\ 3 of Ref.\cite{parolini2016}). The average number of complexes is then given by 
\begin{eqnarray}
\on_{t_1}=\on_{t_2} =\on_{t_B} &=& m {\on_{\delta_1} \on_{\delta_2}   \on_{\gamma} \over (\rho_\ominus L A)^2 }  \Big( 2-{h\over L} \Big)\exp[-\beta \Delta G^0_{\gamma\delta_1\delta_2}]
\nonumber \\
\on_{t_3} &=& m {\on_{\delta_1} \on_{\delta_2}   \on_{\gamma} \over (\rho_\ominus L A)^2 }   \exp[-\beta \Delta G^0_{\gamma\delta_1\delta_2}]
\end{eqnarray}
where $m$ is a multiplicity factor that counts the number of iso-energetic states ($m=5$ in our case \cite{parolini2016}).
The top panel of Fig.\ 
3{\bf d} in the main text reports the amount of three strand complexes {\em per} $\gamma$ strand
\begin{eqnarray}
{ f_3}= {\on_{t_1}+\on_{t_2} + \on_{t_B} + \on_{t_3}  \over N }
\end{eqnarray}
For isolated vesicles only $t_3$ is present and we have
\begin{eqnarray}
\on^0_{t_3} &=& m {\on^0_{\delta_1} \on^0_{\delta_2}   \on^0_{\gamma} \over (\rho_\ominus L A)^2 }   \exp[-\beta \Delta G^0_{\gamma\delta_1\delta_2}]
\, .
\end{eqnarray}

{ By applying} Eq.\ 7 of the main text we can finally calculate the free energy {\em per} $\gamma$ strand of the system 
\begin{eqnarray}
\beta { F}_\mathrm{att}  &=&  
\log {\on_{\gamma} \over N} + \chi_1 \log {\on_{\delta_1}\over \chi_1 N} + \chi_2\log {\on_{\delta_2} \over \chi_2 N} +
{\on_{b_1}\over N} + { \on_{b_2} \over N} +
{\on_{\ell_1}\over N} + { \on_{\ell_2} \over N} 
\nonumber \\
&& \qquad \qquad +2 {\on_{t_1} \over N}+2 { \on_{t_2} \over N}+2 {\on_{t_B} \over N} + 2 {\on_{t_3} \over N} 
\end{eqnarray}
 On the other hand for isolated vesicles we have  
 \begin{eqnarray}
\beta { F}^0_\mathrm{att} (h=\infty) &=& 
\log {\on^0_{\gamma} \over N} + \chi_1 \log {\on^0_{\delta_1}\over \chi_1 N} + \chi_2\log {\on^0_{\delta_2} \over \chi_2 N} +
{ \on^0_{\ell_1} \over N} + { \on^0_{\ell_2} \over N} + 2 { \on^0_{t_3} \over N} 
\end{eqnarray}
 
 Finally in Fig.\ 3{\bf c} of the main text we report the interaction free energy given by $\beta { F}_\mathrm{att}
 -\beta { F}^0_\mathrm{att}$.